# Determination of SATI Instrument Filter Parameters by Processing Interference Images

Atanas Atanassov

*Solar Terrestrial Influences Institute, Bulgarian Academy of Sciences,
Stara Zagora Department, P.O. Box 73, 6000 Stara Zagora, Bulgaria,
At_M_Atanassov@yahoo.com*

**Summary**

This paper presents a method for determination of interference filter parameters such as the effective refraction index $\mu$ and the maximal transmittance wavelength $\lambda_0$ on the basis of image processing of a spectrogram produced by Spectrometer Airglow Temperature Imager instrument by means of data processing. The method employs the radial sections for determination of points from the crests and valleys in the spectrograms. These points are involved in the least square method for determination of the centres and radii of the crests and valleys. The use of the image radial sections allows to determine the maximal number of crests and valleys in the spectrogram. The application of the least square fitting leads to determination of the image centers and radii of the crests and valleys with precision higher than one pixel. The nocturnal course of the filter parameters produced by this method is presented and compared with that of the known ones. The values of the filter parameters thus obtained are closer to the laboratory measured ones.

**Introduction**

The Spectrometer Airglow Temperature Imager (SATI) is a ground-based spatially scanning Fabry-Perot interferometer [1]. The basic principles for the instrument design, operation and data processing were developed about two decades ago by the Space Instrumentation Laboratory at the Centre for Research Earth and Space Science (CRESS) at York University - Canada [2-4]. A version of SATI-3SZ was developed at the Stara Zagora Division of the Solar-Terrestrial Influences Laboratory in collaboration with CRESS laboratory [5].

The measurements of airglow by Bulgarian electrophotometers have been started in 1969 in the Basic Observatory of the Central Laboratory for Space Research, BAS, Stara Zagora city [6,7]. In the first Bulgarian theoretical investigations of airglow a new mechanism of the green and red oxygen lines variations was proposed. This mechanism takes into account the ion chemistry of the D- region under the influence of natural optical emissions. These problems were elaborated in a series of publications of Gogoshev, Serafimov, Velinov et al. [8-13].

**SATI description**

The light comes to the SATI instrument from the direction of an annular sky segment through a conic mirror (Fig. 1). After that a narrow-band interference filter disperses it and a camera objective forms an interference pattern in the focal plane. The image is registered by a Charged Coupled Device (CCD). Every radial section of the interference image represents a spectrum of light which has come from the respective section of the annular segment. The volume emission maximum for $O_2(b^1\Sigma_g^+ - X^3\Sigma_g^-)$ transmissions is emitted at the altitude of the mesopause.

The analysis of this spectrum allows to determine the rotational temperature of the emitting gas. The time variations of the temperature values in different points of the annular segment at the altitude of the mesopause enable the investigation of wave processes. The temperature is determined by comparing the measured spectrum with the so-called synthetic spectra, calculated in advance [2]. The synthetic spectra are series of emission intensity values, calculated for the respective $O_2$ transitions and convoluted with the filter transmittance function. The calculations are made for different temperatures of the mesopause within the range 110÷300°K with a wavelength step, 0.01nm in this case.

The following relation between the incident angle $\Theta$ and the wavelength $\lambda_i$ (for each crest and valley) is obtained from the basic equation of the Fabry-Perrot etalon applied to the SATI instrument [1]

$$\sin\Theta = \mu\sqrt{1 - \left(\frac{\lambda_i}{\lambda_0}\right)^2}$$

Parameters $\mu$ and $\lambda_0$ are the effective refraction index and the maximal transmittance wavelength of the filter, respectively [2]. On the other hand, geometrical considerations (Fig. 1) yield

$$\tan\Theta = \frac{r_i}{f},$$

where $r_i$ is the radius of the respective crest or valley in the image and $f$ is the focal distance of the camera objective. On the basis of the last two relations we obtain

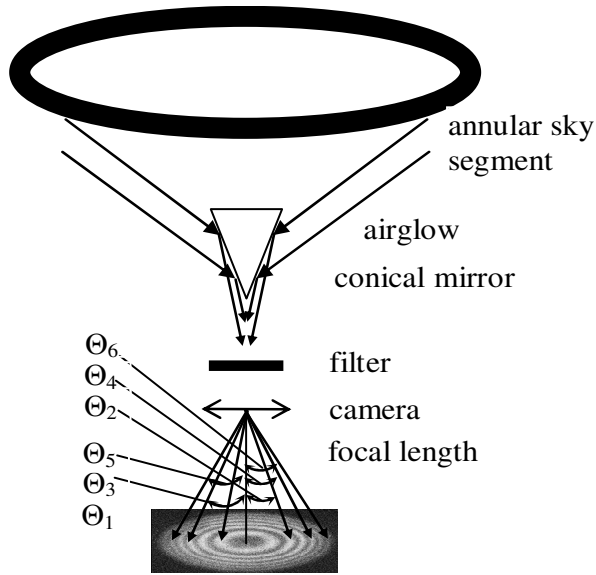

**Fig. 1.** General scheme of SATI instrument

(1) $$\frac{r_i^2}{r_i^2 + f^2} = \mu^2\left(1 - \left(\lambda_i/\lambda_0\right)^2\right), \quad i = \overline{1, m}$$

The symbol *m* in the last formula denotes the number of used crests and valleys. The determination of parameters $\mu$ and $\lambda_0$ is on the basis of (1) by a linear regression based on connection (1) [2]. For this purpose (1) was first transformed into

(2) $\quad \eta_i = b_1 \zeta_i + b_0,$

The following replacements $\eta_i = \lambda_i^2$, $\zeta_i = r_i^2/(r_i^2 + f^2)$, $b_1 = -\lambda_0^2/\mu^2$ and $b_0 = \lambda_0^2$ are made in the last equation. Values $\eta_i$ and $\zeta_i$ correspond to the respective crest or valley and are known. The filter parameters are calculated through the above substitution as $\lambda_0 = \sqrt{b_0}$ and $\mu = \sqrt{-b_0/b_1}$ after finding $b_1$ and $b_0$ from (2).

In order to compare the measured $S^p$ and the synthetic $S^\lambda$ spectra, $S^\lambda$ are transformed from the frequency space into the pixel one on the basis of the general type of transformation

(3) $\quad S_p^\sim(T) = \sum_{\lambda p}^{\lambda p+1} \xi(S_\lambda(T), \mu, \lambda_0)$

Basically, non-analytic function $\xi$ determines the correspondence between the synthetic spectrum and the measured one. To perform this transformation, the correspondence between the wavelength and the pixels of the interference pattern should be known. The wavelength $\lambda_p$ which forms the image at distance $r_p$ (in pixels) from the image center can be presented as

(4) $\quad \lambda_p = \left(\lambda_0^2 \left(1 - \frac{r_p^2}{\mu^2(r_p^2 + f^2)}\right)\right)^{1/2}$

In (4) the focal distance *f* and the distance between the pixels from the image centre are presented in the same units.

The filter parameters $\mu$ and $\lambda_0$ depend on the filter temperature, therefore preventive measures were undertaken in order that they remain relatively constant within a range of <1°C. Furthermore, the filter temperature should be as close as possible to the temperature at which the filter transmittance function is measured, which is used for the convolution of the synthetic spectra. Thus, the maintenance of quasi-constant filter temperature is important to decrease the errors when determining the rotation temperature by means of comparing the measured with the synthetic spectra.

An averaged spectrum is formed in the original algorithms for processing the interferograms produced by the SATI instruments [6]. The values of all pixels at equal distances from the image centre are averaged. The spectrum thus produced is filtered by a mean filter. After that the crests and valleys are searched by equal patterns. The number of found crests and valleys depends on the contrast of every image.

The purpose of this paper is to present part of the algorithms for processing the patterns produced by the SATI instrument. In addition to the algorithms for filter parameters determination, the results from the comparison between these algorithms and the original ones [4, 14] are presented. Prior to applying the algorithm for determination of the filter parameters, every image is processed for removing some pixel defects, for dark-image subtraction and high frequency space _filtering. Some other ways to solve these problems are presented in [15]. They are different from the original ones [4, 14].

### Overview of instrument operation

The operation of the SATI-3SZ instrument is automatic and PC-controlled. The measurements start after astronomical twilight in the beginning of the night and finish before its beginning in the morning, before sunrise and when the Moon is under the horizon. Each spectrogram is obtained with exposure time of 120s. As a result of the measurement sequence, two series of images are accumulated, from which data are extracted.

- $I_{m,n}(t_{im})$, im=1,...,P- spectrogram images
- $D_{m,n}(t_d)$, d=1,...,Q- dark current images

After each measurement sequence, depending on its duration, spectrograms with P number and Q dark images are available. The (m,n) indices determine the location of the pixels in the bi-dimensional image. The time step, when dark current images are obtained, is almost constant. The images, containing spectrograms are also formed with an almost constant time step. After each sequence of eight observation images, a dark current image is taken.

### Determination of radii and centres of crests and valleys in an image

The following approach is applied here to determine the respective radii of crests and valleys. First, an initial approximation of the interference rings centre is selected $O^0(m^0,n^0)$ and on the basis of the image radial sections $S_{k,\varphi}$, the locations of the respective maxima are specified, based on a special procedure [15]. The radial section is defined as

$$S_{k,\varphi} = I'_{k.cos(\varphi),k.sin(\varphi)}; \quad k = 1 \div 128; \quad \varphi = 0 \div 360°; \quad \Delta\varphi = 1°$$

Here $I'$ is a filtered image. The section length depends on the size and place of the image centre on the CCD and its maximum is 128 pixels in our case.

An approach has been developed for search and recognition of maxima, which are not sufficiently distinguished in the presence of residual noise. Every interference maximum possesses own individuality - the central ones are better expressed, while the outer ones sometimes and in some parts of the image might appear smeared and not very clear.

Each interference ring can be searched on the basis of different patterns $P_k$ depending on the image contrast. Every pattern $P_k(\{b_i\}, i = 1, L)$ is an arranged binary multitude $\{b_1, b_2, \ldots, b_L\}$, as each element describes a relation between two adjacent points (increase or decrease) from the section. Since the search procedure is not always productive, it is necessary to reduce the criteria and more or less recognize an interference maximum in the space of relations with a suitable pattern. This approach provides the opportunity to increase the number of points on the outer crests and valleys found. This is of major importance for the interference images with weaker contrast, obtained with not very strong useful signal.

### Least square procedure for centre and radii determination

After determining the coordinates of the interference maxima and minima within the frames of the radial sections it is necessary to allocate them according to the interference rings. This is required because the maxima of all rings are not always found. In such cases the problem is solved on the basis of the analysis of the histogram for distribution of all determined maxima [15].

Having enough available points for each ring (crest or valley), the Least Square Fitting (LSF) can be applied to determine the centres and their radii. A multitude of points $\{(x_k, y_k)\}_{k, l=1,N}$ for each $k^{th}$ ring is obtained, which should satisfy the condition to lie as close as possible to a circle with radius $r$ and centre coordinates $(a, b)$, that is $(x-a)^2 + (y-b)^2 = r^2$ have to be fulfilled. Since the determined points do not lie exactly on the circle, in order to define that circle, it is necessary to minimize the following functional:

$$(5) \quad E(a,b,r) = \sum_{l=1}^{N}(L_l - r)^2, \quad L_l = \sqrt{(x_l - a)^2 + (y_l - b)^2}$$

The functional $E(a,b,r)$ is minimized by an approach, which does not lead to solving a non-linear LSF problem. An iteration procedure is applied [16], which is a convergence and yields very good results.

$$(6a) \quad a = \bar{x} + \bar{L}\bar{L}_a, \quad b = \bar{y} + \bar{L}\bar{L}_b,$$

$$(6b) \quad \bar{x} = \frac{1}{N}\sum_{l=1}^{N} x_l, \quad \bar{y} = \frac{1}{N}\sum_{l=1}^{N} y_l,$$

$$(6c) \quad \bar{L} = \frac{1}{N}\sum_{l=1}^{N} L_l, \quad \bar{L}_a = \frac{1}{N}\sum_{l=1}^{N}\frac{a - x_l}{L_l}, \quad \bar{L}_b = \frac{1}{N}\sum\frac{b - y_l}{L_l}$$

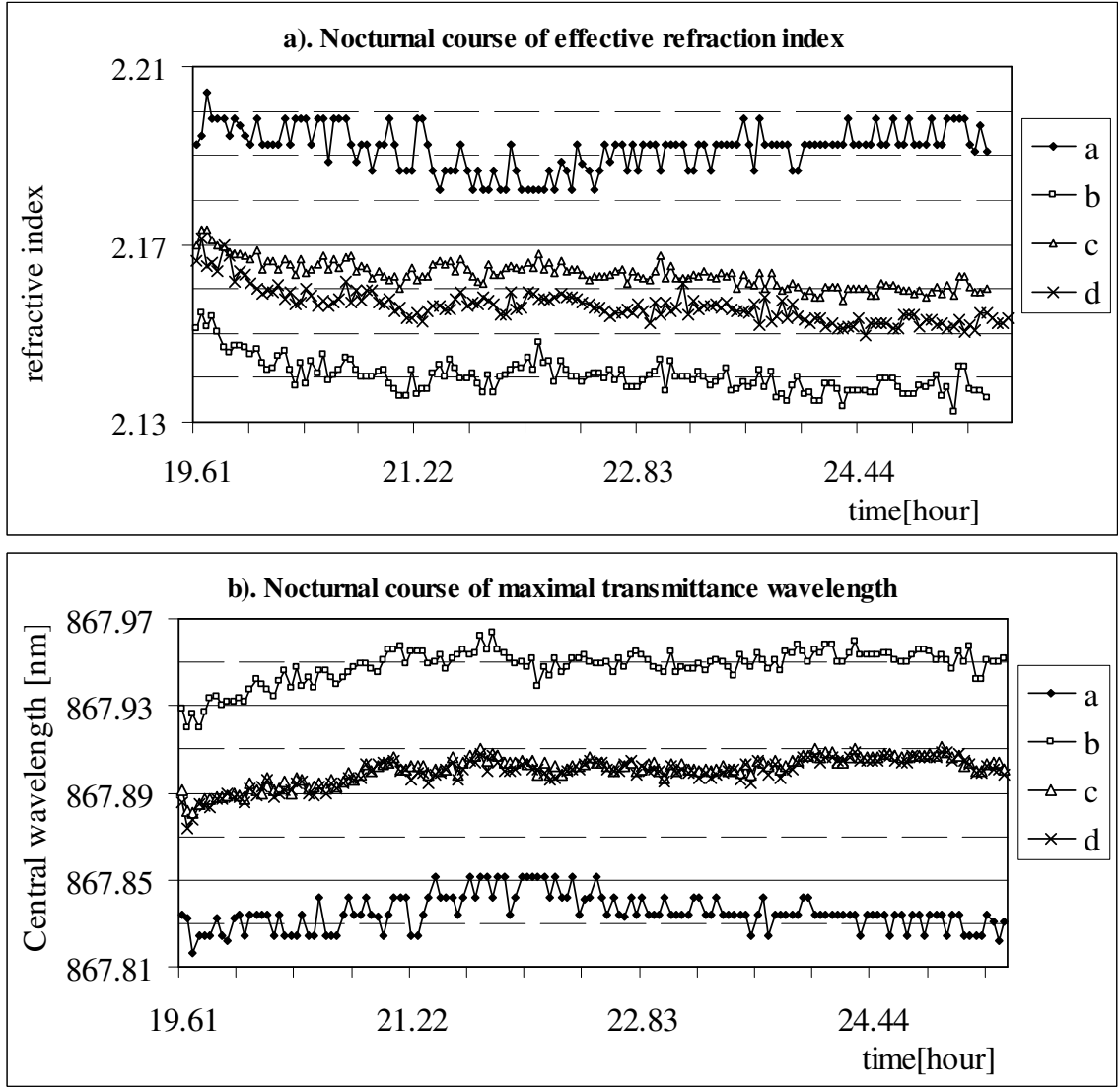

**Fig. 2.** Nocturnal course of filter parameters; series *a* are produced by original [14] and series *b*, *c* and *d* by presented algorithm respectively.

Before starting the iteration process, as an initial approximation of the centre coordinates it is assumed that $a_0 = \bar{x}$, $b_0 = \bar{y}$. After completing the iteration process and defining the centre coordinates its radius is determined as $r = \frac{1}{N}\sum_{l=1}^{N} L_l$. The entire process of defining the ring parameters is applied iteratively for their improvement. This process requires a good initial approximation of the image centre and it is quickly convergent. The proof for the process convergence is illustrated numerically [15]. The next step is the regression problem to determine the filter parameters µ and $\lambda_0$.

### Comparison of the new and the original algorithms

The nocturnal course of the above filter parameters is shown in Fig. A, B. The series denoted with symbol *a* is obtained by the original algorithm. It has produced eight crests and valleys. Series *b* is calculated with the proposed algorithm on the basis of eight crests and valleys. The possibilities of the new approach to find all eleven crests and valleys allow to calculate series *c*. The last series *d* is determined with six crests only.

The differences between the results of the two approaches are due to several reasons. The first of them is connected with the precision when determining the centers of the crests and valleys, on the one hand, and of their radii, on the other. The accuracy when applying the original approach is restricted to integer pixels both for the centers and for the radii. The proposed approach employs real numbers to present the values of the centers and radii of the interference rings. Moreover, the use of variable types and length patterns and the recognition strategy for searching the maxima and minima in radial image sections lead to determination of the maximal number of crests and valleys.

In the first place, the differences between the two approaches by using four crests and four valleys (maximal possible for the original approach applied to our data) for one ordinary séance are ∆µ~.05 and $\Delta\lambda_0$~0.012nm. The use of the overall possibilities of the presented approach yields results which are very close to the laboratory measured values (µ=2.1551, $\lambda_0$=868.05nm)

### Conclusion and future work

The first part of algorithms of the approach presented in this paper is connected with determination of points which lie on crests and valleys. The application of recognition algorithms instead of searching ones allows to determine the largest possible number of points for the outer interference rings. Thus, the centers and radii of all eleven crests and valleys are determined with high precision even in cases of a low signal. The second part of algorithms demonstrates the application of LS fitting for determination of the centers and radii of the crests and valleys. The values of the filter parameters calculated by our approach are closer to those determined in laboratory conditions as compared with the original ones.

The calculation of filter parameters is an important stage in the image processing whose final goal is the determination of mean and sectors rotational temperatures. The development of the presented approach is a major stage in the development of new algorithms whose purpose is to improve the results in the calculation of the rotational temperature. It would be interesting as a next step in the investigation of the algorithms for determination of filter parameters to compare the results of the synthetic spectra transformation from the frequency space into the image space where the latter can be compared to the measured spectra. This is important in order to evaluate the possibilities of the two approaches for precise determination of the rotational temperature. A forthcoming stage would be the preparation of a numerical experiment with a computer generated image on the basis of preset values of the filter parameters and verification of the algorithm possibilities to confirm these values.

The development of new processing algorithms is aimed at improving the accuracy of the final results- rotation temperature determination. The synthetic spectra, transformed into the pixel space by the two approaches can be compared as a next step in the investigation of their precision. A numerical experiment is under preparation for computer generation of interference images by selecting filter parameter values (as a forward problem). A test application of the described procedures on these artificially prepared images (as an inverse problem) will show their possibilities to approach the initial values of the selected filter parameters.

**Acknowledgements.** We acknowledge Corresponding Member P. I. Y. Velinov for the discussions and for the help in the preparation of this paper; and also Dr. Y. Tassev and Dr. L. Mateev for the technical support.